\documentclass[aps,prl,superscriptaddress,twocolumn]{revtex4-2}
\usepackage{graphicx} 
\usepackage{amsmath,amssymb}
\usepackage{bbm, dsfont,braket}
\usepackage{hyperref}
\newcommand{\Tr}{\mbox{Tr}\,}
\newcommand{\e}{\mathrm{e}}
\newcommand{\im}{\mathrm{i}}

\usepackage{color}
                
\newcommand{\js}[1]{{\color{black}#1}}
\newcommand{\ds}[1]{{\color{black}#1}}
\newcommand{\blue}[1]{{\color{black}#1}}
\newcommand{\ii}{\text{i}}

\allowdisplaybreaks[4]

\begin{document}

\title{\ds{Parastatistics in Interacting Periodic Chains Revealed by Peierls Phase Twists and Shifted Conformal Towers}}
\author{Dirk Schuricht}
\affiliation{Institute for Theoretical Physics, Center for Extreme Matter and Emergent Phenomena, Utrecht University, Leuvenlaan 4, 3584 CE Utrecht, The Netherlands}
\author{Jesko Sirker}
\affiliation{Department of Physics and Astronomy and Manitoba
Quantum Institute, University of Manitoba, Winnipeg, Canada R3T 2N2}
\date{\today}
\begin{abstract}
We consider interacting paraparticle chains with a constant $R$-matrix where the Hamiltonian sums over the internal degrees (flavors) of the paraparticles. For such flavor-blind Hamiltonians, we show a general factorization of the Hilbert space into occupation and flavor parts with the Hamiltonian acting non-trivially only on the former. For open boundaries, the spectrum therefore coincides with that of the occupation Hamiltonian $H_{\rm occ}$ with the flavor part merely adding degeneracies. For periodic boundaries, a cyclic reordering of the flavors leads to a separation of $H_{\rm occ}$ into flux sectors at fixed particle number, thus making the parastatistics directly observable in the energy spectrum. For important exemplary cases, $H_{\rm occ}$ reduces to the XXZ chain with flux, allowing for an exact solution. In the gapless regime, this solution shows flux-shifted $c=1$ conformal towers in the low-energy spectrum and a temperature-dependent chemical potential in the bulk thermodynamics.
\end{abstract}
\maketitle

\emph{Introduction}---When considering identical quantum particles, the distinction between bosons and fermions is of fundamental importance. For example, it manifests itself in the formation of Bose--Einstein condensates in ultracold atomic gases and Fermi surfaces in solids, both of which in turn dictate the macroscopic physical properties. At a more technical level, bosons and fermions are distinguished by the behavior of the many-body wave functions under particle permutations, which can be encoded in the commutation relations of the respective creation and annihilation operators \cite{FetterWalecka}. 

Given the profound consequences of different quantum statistics, there have been various efforts to go beyond bosons and fermions. To give just a few examples, in two dimensions one can consider braiding instead of permutations~\cite{LeinaasMyrheim77},  leading to so-called anyons~\cite{Wilczek82,Nayak08}, which emerge as quasiparticles in fractional quantum Hall states~\cite{Arovas-84,Halperin84,Toppan24}. Anyonic statistics is also displayed by $\mathbb{Z}_p$-parafermions~\cite{FradkinKadanoff80, Fendley12}---generalizations of Majorana fermions in Potts/clock models---which are suspected to exist in quantum Hall/superconductor hybrid systems and topological insulators~\cite{Mong-14,AliceaFendley16}. Furthermore, building on parafermions one can construct so-called Fock parafermions~\cite{CobaneraOrtiz14,Rossini-19,MWS20}, which show a generalized Pauli principle in the sense that single-particle levels can be occupied by at most \ds{$p-1$} particles. In fact, exclusion statistics even with fractional statistical parameters can be defined in any dimension~\cite{Haldane91prl2,Wu94} and is realized by spinons in antiferromagnetic spin chains~\cite{Haldane91prl1} and conformal field theories~\cite{Schoutens97}. However, all generalizations mentioned above appear only in strongly correlated systems and thus are not amenable to simple single-particle descriptions. A notable exception is the Baxter--Fendley model~\cite{Baxter89,Fendley14}, whose many-particle spectrum is built from single-particle energies of $\mathbb{Z}_p$-parafermions. However, the model is non-Hermitian, thus obscuring its quantum mechanical interpretation. 

Very recently, Wang and Hazzard~\cite{WangHazzard25} introduced a parastatistics based on non-trivial bilinear relations between the second-quantized operators acting on internal flavors,
in contrast to Green-type trilinear algebras \cite{Green53,GreenbergMessiah65}, which complicate thermodynamic derivations \cite{StoilovaJeugt20}. For the newly introduced paraparticles, Wang and Hazzard studied the generalized exclusion and exchange statistics and derived the exact energy spectra of certain flavor-blind bilinear Hamiltonians by relating them to specific quantum spin chains with open boundary conditions (OBC). In this sense, for flavor-blind bilinear chains with OBC, the Hamiltonians realize models of free paraparticles whose many-body spectra are built from single-particle energies with a generalized exclusion principle encoded in the mode \ds{multiplicities $d_n$~\cite{WangHazzard25,LiLiu,Dakic}.}

In this Letter, we go beyond free paraparticle chains with OBC by \js{providing the first systematic study of interacting paraparticle chains with periodic boundary conditions (PBC).}
In contrast to the OBC case, the single-particle energies are already modified by a Peierls twist due to the parastatistics for periodic boundaries. \js{Importantly, this twist is not imposed externally but emerges solely from the statistics encoded in the quadratic 
R-algebra.} Our main general results are: (i) a generic factorization theorem 
for the Hilbert space of flavor-blind Hamiltonians, (ii) an explicit derivation of the differences between OBC and PBC: with OBC the \js{flavors only contribute degeneracies} while with PBC the cyclic permutation of flavors induces a Peierls twist in $H_{\rm occ}$, splitting the energy spectrum into flux sectors, and (iii) an exact formula for the dimension of the flavor subspaces in terms of character projectors of the cyclic group. For the specific case that the single-mode occupations are given by $d_0=1$, $d_1=m$, and $d_n=0$ for $n\geq 2$, we show, furthermore (iv) that the occupation sector for paraparticles with nearest-neighbor interaction is the Bethe-ansatz-solvable XXZ chain with flux. This exact solution shows (v) that for PBC the low-energy spectrum consists of conformal towers with a shift between persistent current branches, and (vi) that the bulk thermodynamics shows two signatures of the paraparticle character of the constituent particles: a zero-temperature residual entropy and a temperature-dependent chemical potential. \js{Together, these results establish interacting periodic chains as concrete benchmark systems in which R-parastatistics leads to directly observable many-body signatures.}

\emph{General setup}---We use the second-quantized formulation of parastatistics recently introduced in Ref.~\cite{WangHazzard25}. Specifically, we consider a one-dimensional chain of $L$ lattice sites, at each of which we define operators $\psi_{i,a}^\pm$, $i=1,\ldots,L$, that create or annihilate a particle with internal flavor $a=1,\ldots, F$. The parastatistics is encoded in the commutation relations 
\begin{eqnarray}
\label{comm}
&&\hspace{-4mm}\psi_{i,a}^+\psi_{j,b}^+=
\sum_{cd} R^{cd}_{ab}\psi_{j,c}^+\psi_{i,d}^+,\quad
\psi_{i,a}^-\psi_{j,b}^-=
\sum_{cd} R^{ba}_{dc}\psi_{j,c}^-\psi_{i,d}^-,\nonumber\\
&&\hspace{-4mm}\psi^-_{i,a}\psi^+_{j,b} =\sum_{cd} R^{ac}_{bd} \psi^+_{j,c}\psi^-_{i,d}+\delta_{ab}\delta_{ij},\label{Para}
\end{eqnarray}
where $R_{ab}^{cd}$ are \js{$F^2\times F^2$} matrices satisfying $\sum_{\sigma\tau}R_{ab}^{\sigma\tau}R_{\sigma\tau}^{cd}=\delta_a^c\delta_b^d$ and $\sum_{\sigma\tau\kappa}R_{ab}^{\sigma\tau}R_{\tau c}^{\kappa u}R_{\sigma\kappa}^{de}=\sum_{\sigma\tau\kappa} R_{bc}^{\sigma\tau}R_{a\sigma}^{d\kappa}R_{\kappa\tau}^{eu}$. \js{In the following we also assume unitarity, $\sum_{\sigma\tau}R_{ab}^{\sigma\tau}(R_{cd}^{\sigma\tau})^*=\delta_{ac}\delta_{bd}$, so that $(\psi_{i,a}^+)^\dagger=\psi_{i,a}^-$ and $H$ is Hermitian with a real spectrum and a standard Gibbs ensemble.} We note that the $R$-matrix defined in \eqref{Para} is known as the permuted $R$-matrix in the literature of integrable systems~\cite{SamajBajnok13}, with the latter relation being the constant Yang--Baxter equation. In the special case $R_{ab}^{cd}=\pm\delta_a^d\delta_b^c$ the relations \eqref{Para} simplify to bosons and fermions with $F$ internal degrees of freedom. The constant YBE above defines braid generators $B_i$ built from $R$ acting on neighboring flavor spaces and satisfying the braid relations. The first quadratic condition below Eq.~\eqref{comm} implies $R^2=1$, so that $B_i^2=1$ and the braid representation reduces to a representation of the symmetric group. Exchange eigenvalues are therefore restricted to $\pm1$, and no braid-type exchange statistics can arise within this class. The allowed statistics thus consists of bosonic and fermionic exchange channels which can be coupled by local constraints, leading to generalized exclusion rules and flavor-dependent degeneracies beyond simple independent bosons and fermions (see End Matter). For $F=2$, all constant solutions of the YBE have been classified by Hietarinta~\cite{Hietarinta}, and we analyze the corresponding generalized statistics in the \ds{Supplemental Material~\cite{supplement}.}

Using the second quantized operators above, we introduce the OBC or PBC Hamiltonian 
\begin{equation}
\label{Ham}
    H=J\sum_{i,a} (\psi^+_{i,a}\psi^-_{i+1,a}+\mathrm{h.c.})+\sum_{i<j} V_{|i-j|} n_i n_j -\mu \sum_i n_i,
\end{equation}
where $n_i=\sum_a n_{i,a}$ and $n_{i,a}=\psi^+_{i,a}\psi^-_{i,a}$ are the total and flavor particle densities, respectively. We note that in analogy to Ref.~\cite{WangHazzard25}, a sum over the internal flavor degrees is performed, making the Hamiltonian flavor-blind. In contrast to Ref.~\cite{WangHazzard25}, we do include an explicit interaction term which is allowed to be long range for the general discussion below. In the specific examples considered at the end of this letter, the interaction will be limited to nearest neighbors only. For a single mode, we denote the number of possible states with $n$ particles by $d_n$ \cite{WangHazzardSM}. \blue{The factorization theorem and the OBC and PBC degeneracy rules discussed below are general consequences of the quadratic $R$ algebra together with the flavor-blind form of the Hamiltonian; they do not rely on integrability.}

\emph{Hilbert space structure}---The first important point is that for \ds{flavor-blind Hamiltonians} of the form \eqref{Ham}, with operators satisfying the algebra \eqref{comm}, the Hilbert space factorizes.\\
{\bf Factorization theorem:} \js{At each lattice site $i$, the local Hilbert space admits a decomposition using the basis states $|n_i\rangle$ of the local density operator $n_i=\sum_a\psi^+_{i,a}\psi^-_{i,a}$ as
\begin{equation}
\label{Hilbert}
\mathcal{H}_i=\bigoplus_{n=0}^{n_{\rm max}} \mbox{span}\{|n_i\rangle\} \otimes \mathcal{F}_n
\end{equation}
where the local flavor space dimension, $\mbox{dim}\,\mathcal{F}_n=d_n$, is determined by the algebra \eqref{comm}. Because the Hamiltonian \eqref{Ham} sums over the internal flavor index, its action on the internal indices is completely fixed by the quadratic algebra and does not introduce additional flavor dependence; in particular, within each flavor sector, defined by the flavor symmetry algebra commuting with the Hamiltonian, it acts trivially on the corresponding \ds{flavor} multiplicity space. The full Hilbert space therefore separates into an occupation sector and a flavor sector which is merely a spectator and contributes only degeneracies. A proof of this theorem is given in the End Matter. For OBC, the linear ordering of sites fixes a canonical identification of flavor spaces across different occupation configurations, so} the degeneracy is simply the dimension of the flavor space compatible with the occupation number eigenstate. A general state at fixed particle number $N$ is given by $|\Psi\rangle_N = \sum_s a_s |\{n_i\}_s\rangle \otimes |\phi_{\rm fl}\rangle$ with $\sum_i n_i=N$ for each $s$. Each occupation number configuration $|\{n_i\}\rangle$ has a compatible flavor space $\mathcal{F}(\{n_i\})=\bigotimes_i \mathcal{F}_{n_i}$.\\
{\bf OBC degeneracy rule:} The dimension of the flavor space \js{compatible with all occupation configurations entering the superposition $|\Psi\rangle_N$} is $D=\mbox{dim}\left[\bigcap_{a_s\neq 0} \mathcal{F}(\{n_i\}_s)\right]$. 

To make this concrete, we consider example 3 in Ref.~\onlinecite{WangHazzard25} where $R_{ab}^{cd}=-\delta_a^c\delta_b^d$ and $d_0=1$, $d_1=m$ (which equals $F$ in this example), $d_n=0$ for $n\geq 2$ (thus $n_{\rm max}=1$). In this case the flavor space for $N$ particles is always the same, $\mathcal{F}=\left(\mathcal{F}_1\right)^{\otimes N}$ with $\mbox{dim}\,\mathcal{F}_1=m$, independent of the concrete particle configuration $|\{n_i\}\rangle$ implying $D=m^N$. If, on the other hand, $d_2\neq 0$ (e.g.~$d_0=1,\, d_1=m,\, d_2=1$) then superpositions like $|\cdots 1,1\cdots\rangle+|\cdots 2,0\cdots\rangle$ are possible, forcing the flavor factor on the doubly occupied site to be the 1-dimensional $\mathcal{F}_2$ and thus reducing $D$. If the spectrum of the Hamiltonian acting on the occupation numbers is known---for example, if $H$ is non-interacting or integrable---then the spectrum of the paraparticles is simply constructed by taking the additional degeneracy $D$ of each occupation eigenstate due to the flavor sector into account. However, while the degeneracies are related to the single-mode occupation numbers $d_n$, the exchange statistics of the paraparticles is hidden for OBC because flavors are never commuted. 

\blue{This all changes for PBC, where the flavor sector, which for OBC only contributes degeneracies, becomes directly visible in the spectrum. We can then wrap a flavor around the chain, thus obtaining cyclic permutations in the order of the flavors.} To understand how the eigenstates of the paraparticle Hamiltonian can be constructed from an occupation number Hamiltonian plus flavor degeneracies, one then has to study some basic properties of the cyclic group first. We define $N_n$ as the number of particles occupying modes with degeneracy $d_n$. Then $N=\sum_n N_n=\sum_i n_i$ is the total particle number. We define, furthermore, $M=M(\{N_n\})$ as the total number of flavors which actually does get cyclically permuted under PBC. It is important to note that the length of the flavor string $M$ is, in general, not equal to the number of particles $N$. Consider, for example, the case $d_0=1$, $d_1=2$, $d_2=1$, and $d_n=0$ for $n>2$. In this case $M=N_1$, where $N_1$ is the number of singly occupied modes because the vacuum and the doubly occupied modes do not have a flavor index. 

Assume that we have a flavor state $|\phi_{\rm fl}\rangle =|\alpha_1\cdots\alpha_M\rangle$ where $\alpha_i$ is a flavor label. The cyclic permutation operator $C$ then acts on the state as $C|\alpha_1\cdots\alpha_M\rangle =|\alpha_2\cdots\alpha_M\alpha_1\rangle$ implying that $C^M=\mathbbm{1}$. The eigenvalues of $C$ are therefore given by $\lambda_q=\exp(\im\gamma_q)$ with $\gamma_q=2\pi q/M$ and $q=0,\cdots,M-1$. A simple example is the case of two flavors $\{a,b\}$ with $M=2$. Then the flavor space is $4$-dimensional and splits into a $3$-dimensional symmetric eigenspace $\{|aa\rangle,|bb\rangle,(|ab\rangle+|ba\rangle)/\sqrt{2}\}$ with $\lambda_0=+1$ and a $1$-dimensional anti-symmetric eigenspace $(|ab\rangle -|ba\rangle)/\sqrt{2}$ with $\lambda_1=-1$. Next, we consider an $N$-particle eigenstate of paraparticles $|\Psi\rangle=|\Psi_{\rm occ}(x_1,\cdots,x_N)\rangle\otimes |\phi_{\rm fl}(\alpha_1,\cdots,\alpha_M)\rangle$ where $x_1<\cdots<x_N$ are the positions of the particles and $\alpha_i$ the flavor labels. The flavor part has to be an eigenfunction of the cyclic permutation operator $C$ and the total eigenfunction $|\Psi\rangle$ has to be single valued and invariant under \ds{a cyclic reordering}. If we assume that $x_1$ is the position of a particle with flavor label $\alpha_1$ then
\begin{eqnarray}
    \label{PBC}
    &&\!\!\!\!\!|\Psi_{\rm occ}(x_2,\cdots,x_N,x_1)\rangle\otimes |\phi_{\rm fl}(\alpha_2,\cdots,\alpha_M,\alpha_1)\rangle \\
    &&= \e^{-\im\delta}|\Psi_{\rm occ}(x_1,\cdots,x_N)\rangle\otimes \e^{\im\gamma_q}|\phi_{\rm fl}(\alpha_1,\cdots,\alpha_M)\rangle \nonumber
\end{eqnarray}
with $\delta\equiv\gamma_q$. That is, the occupation number wavefunction $|\Psi_{\rm occ}\rangle$ picks up a phase which is equal and opposite to the phase picked up by the flavor part. The eigenspaces of the occupation number Hamiltonian for PBC thus separate into spaces with Peierls phases $\gamma_q$ which is equivalent to saying that there is now a flux penetrating the ring. As a consequence, the paraparticle statistics---which is responsible for the phases $\gamma_q$---is directly reflected in the eigenspace structure and energies of the occupation number Hamiltonian $H_{\rm occ}$ for PBC. 

The remaining task is to determine the dimension of the flavor eigenspace for fixed $M$ and $q$. We can define the projector onto the eigenspace with eigenvalue $\lambda_q$ as $P_q=M^{-1}\sum_{r=0}^{M-1}\exp(-2\pi \im qr/M) C^r$ where $C$ is the cyclic permutation operator \cite{Serre}. If $C$ acts on flavors in the local flavor space with degeneracy $d_n$ via some representation $\rho_n(C)$ then we define the corresponding character label as $\chi_n(C^r)=\Tr\rho_n(C^r)$. For the flavor-blind Hamiltonians considered here, we have $\rho_n=\mathbbm{1}_{\mathcal{F}_n}$, i.e., the action on the flavor space is trivial and $\chi_n(C^r)=d_n$. We note that one can, in principle, also consider solutions of the constant Yang--Baxter equation which lead to non-trivial actions when cyclically permuting a flavor. Next, we want to calculate $\Tr(C^r)$. To do so, let us define $M_n$ as the number of flavors in the string of type $n$ with $\sum_n M_n = M$. The cyclic permutation operator $C^r$ splits the total of $M$ flavors into $g=\mbox{gcd}(M,r)$ independent groups each of length $\ell=M/g$. This follows from demanding that $C^r$ is the identity on the flavor state implying $\alpha_i=\alpha_{i+r \, (\mathrm{mod}\, M)}$. Furthermore, we must be able to divide the $g$ groups into $c_n$ groups of type $n$ which requires \ds{$c_n=M_n/\ell\in\mathbbm{N}_0$} and $\sum_n c_n = g$. If any $M_n$ is not divisible by $\ell$ then $\Tr(C^r)=0$ and no valid state with this combination of flavors $\{M_n\}$ exists. If all are divisible then there are $g!/\prod_n c_n!$ ways to choose $c_n$ groups out of the $g$ groups and for each group there is then a degeneracy $[\chi_n(C^r)]^{c_n}=d_n^{c_n}$ in the flavor-blind case.\\
{\bf PBC degeneracy rule:} The dimension of the flavor space for fixed $M$ and $q$ is given by
\begin{eqnarray}
    \label{dim}
    &&\mbox{dim}\,\mathcal{H}^{\rm fl}_{M,q} = \Tr P_q =\frac{1}{M}\sum_{r=0}^{M-1}\e^{-2\pi \im qr/M} \Tr (C^r) \nonumber \\
    &&= \left\{\begin{matrix}
    \displaystyle
    \frac{1}{M}\sum_{r=0}^{M-1}\e^{\frac{-2\pi \im qr}{M}} \frac{g!}{\prod_n c_n!}\prod_n d_n^{c_n}, &  \, c_n=
    \frac{M_n}{\ell} \in \ds{\mathbbm{N}_0},\quad\; \\*[0.5cm]
    \ds{0,} & \mathrm{otherwise.}
    \end{matrix}
    \right. 
\end{eqnarray}

\emph{Example}---To demonstrate these general results, we consider from now on example 3 of Ref.~\cite{WangHazzard25} with $d_0=1$, $d_1=m$, and $d_n=0$ for $n\geq 2$. In this hardcore case, there is only one type of flavor which is associated with the single occupancy of a mode and every particle carries one of the $m$ possible flavor labels implying $M=M_1=N$. A natural realization is the $SU(m)$ Hubbard model with on-site interaction $U \to\infty$ and a nearest-neighbor interaction $V$. For the $SU(2)$ case without $V$ a mapping to free spinless fermions with a twist has already been discussed earlier in the literature \cite{CaspersIske89,OgataShiba90,Schadschneider95}  but without any reference to parastatistics and in a model-specific manner, whereas Eq.~\eqref{dim} is completely general and relies on the properties of the cyclic group only. In addition to the hopping terms and chemical potential considered in Ref.~\cite{WangHazzard25}, we include the aforementioned nearest-neighbor density-density interaction with strength $V$. The paraparticle chain can then be mapped onto a spin chain
\begin{equation}
    \label{Ham_spin}
    H=J\sum_{i,\sigma} \big( S^+_{i\sigma}S^-_{i+1\sigma}+{\mathrm h.c.}\big) +V \sum_i n_i n_{i+1}-\mu \sum_i n_i
\end{equation}
with $n_i=\sum_\sigma S^+_{i\sigma}S^-_{i\sigma}$ where $i$ denotes the sites of the lattice and $\sigma=1,\cdots,m$ is the flavor index. Note that in order to obtain an exact solution, we have chosen the hopping $J$, chemical potential $\mu$, and interaction $V$ to be site independent. The following separation into occupation number and flavor part, however, also holds for site-dependent parameters. For this particular model, the separation into these two parts is achieved easily by embedding the Hilbert space into a larger one, $\mathcal{H}\subset\mathcal{H}^{\rm occ}\otimes\mathcal{H}^{\rm fl}$. Here the local occupation Hilbert space is two-dimensional $H^{\rm occ}_i=\mbox{span}(|0\rangle,|1\rangle)$ while the local flavor space is $m$-dimensional, $\mathcal{H}_i^{\rm fl}=\mbox{span}(|s_1\rangle,\cdots,|s_m\rangle)$. To achieve the embedding we simply identify $|0\rangle \equiv |0,s_1\rangle$. The Hamiltonian then becomes
\begin{equation}
    \label{XXZ1}
    H=\bigoplus_{N=0}^L\bigoplus_{q=0}^{N-1} H^{\rm XXZ}(N,q)\otimes\mathbbm{1}_{N,q}
\end{equation}
where the XXZ Hamiltonian, after a Jordan--Wigner transform, is given by
\begin{eqnarray}
&& H^{\rm XXZ}(N,q)= -J\sum_{i=1}^{L-1} (c^\dagger_i c_{i+1}+{\mathrm h.c.})-\mu\sum_{i=1}^L n_i \\
&& +V\sum_{i=1}^{L-1\, (L)} n_in_{i+1} 
\ds{+ J(-1)^N(\e^{i\gamma_q(N)}c^\dagger_L c_1 + \e^{-i\gamma_q(N)}c^\dagger_1 c_L) }\nonumber
\end{eqnarray}
\js{for OBC (PBC)} with $N$ being the particle number. In a block with $N$ fixed, the chemical potential only contributes a constant but does not affect the eigenvectors. The last term is only present for PBC in which case the Peierls phase is $\gamma_q(N)=2\pi q/N$ which can also be distributed uniformly between all bonds. For OBC there is thus no separation into different flux sectors and the dimension of the flavor sector for fixed $N$---which determines the degeneracy of each XXZ eigenvalue---is simply $D=m^N$. For PBC, we can use the general degeneracy formula \eqref{dim} with $c_1=g=\mbox{gcd}(N,r)$ and $d_1=m$ leading to 
\begin{equation}
    \label{dim2}
    \mbox{dim}\,\mathcal{H}^{\rm fl}_{N,q} =\frac{1}{N}\sum_{r=0}^{N-1}\e^{-2\pi \im qr/N} m^g = \frac{1}{N}\sum_{d|N} m^d \,\mathcal{R}_{N/d}(q),  
\end{equation}
where $\mathcal{R}_n(q)$ is Ramanujan's sum \cite{Ramanujan,HardyWright}. To summarize, for PBC the statistics of the paraparticles manifests itself directly in a Peierls phase. Understanding the structure of the Hilbert space furthermore allows \ds{one} to obtain the full paraparticle eigenspectrum from the eigenvalues of the XXZ chain with a Peierls twist for all allowed values of $N,q$ and each of these eigenvalues has a degeneracy equal to $\mbox{dim}\,\mathcal{H}^{\rm fl}_{N,q}$.

While the many-body eigenenergies in the interacting case can be obtained by the Bethe ansatz---or by numerical methods for cases where the occupation number Hamiltonian is not integrable---they can be constructed in the non-interacting case, $V=0$, from the single-particle eigenenergies alone. For OBC, these energies are $\varepsilon_k = -2J\cos k_r-\mu$
with $k_r=\frac{\pi r}{L+1}$, and $r=1,\cdots,L$. The many-body energies are then given by $E(\{n_k\})=\sum_k n_k \varepsilon_k$ with $n_k\in\{0,1\}$, $\sum_k n_k =N$, $N=1,\cdots, L$ and each energy with $N$ particles has flavor degeneracy $m^N$. For PBC, on the other hand, the single-particle eigenvalues are $\varepsilon_k(N,q) = -2J\cos k_{r,q}-\mu$ with \ds{$k_{r,q}=\frac{2\pi r+\gamma_q+\delta_N}{L}$, $r=0,\cdots,L-1$, $q=0,\cdots,N-1$, $\delta_N=0$ for $N$ odd and $\delta_N=\pi$ for $N$ even.} The many-body eigenstates are obtained by going through all allowed values of $N,q$ and in each case constructing all possible eigenvalues $E(\{n_k\})=\sum_k n_k\varepsilon_k(N,q)$ with $\sum_k n_k=N$ and with each one of them having an additional degeneracy of $\mbox{dim}\,\mathcal{H}^{\rm fl}_{N,q}$, see Eq.~\eqref{dim2}.

\emph{Bosonization}---For the flavor-blind Hamiltonians \eqref{Ham} investigated here, the non-trivial part of the low-energy physics is entirely determined by the occupation number part of the Hamiltonian. This part can often be described by ordinary fermions, making it possible to classify the universal behavior using standard techniques. To be concrete, we continue with our example where the occupation number part of the Hamiltonian with PBC is an XXZ chain with a Peierls phase. For $|V/J|<2$, every sector with $0<N<L$ is gapless and described by a conformal field theory with central charge $c=1$.\\ 
{\bf Shifted conformal towers:} The finite-size spectrum in the sector with $N,q$ fixed is given by \cite{Cazalilla,Giamarchi,Bloete-86,Affleck86}
\begin{eqnarray}
    \label{spectrum}
    E(L;N,q)-e_\infty L
&=& -\frac{\pi c v}{6L}
+ \frac{2\pi v}{L}\sum_{n=1}^{\infty} n\,(N_n^{R}+N_n^{L}) \nonumber \\*
&+& \frac{2\pi v K}{L}\min_{J\in \mathbbm{Z}}\left(J-\frac{q}{N}\right)^2 \, .
\end{eqnarray}
Here $e_\infty$ is the energy per site in the thermodynamic limit, $L\to\infty$. The first term on the r.h.s.~is the well known universal finite-size correction with central charge $c=1$ and $v$ the velocity of the excitations \cite{Bloete-86,Affleck86}. The second term is the oscillator contribution, with $N_n^{R/L}$ counting the occupation of the $n$-th mode for right/left-movers, respectively. The parastatistics enters through the third term. In the low-energy effective theory, the Peierls phase $\gamma_q(N)=2\pi q/N$ leads to a persistent current \blue{$I(q)=-\frac{2v K}{L}(J-q/N)$} \ds{with $J=0$ if $q<N/2$ and $J=1$ if $q >N/2$.} If $q=N/2$, then we have a degeneracy corresponding to equal and opposite persistent currents. In the finite-size spectrum the presence of these persistent currents means that we have conformal towers shifted by $\Delta E=\frac{2\pi v K}{L}\min\left( \frac{q}{N}, 1 - \frac{q}{N} \right)^2 $ (for $q$ and $N-q$ the shift is the same, corresponding to opposite persistent currents) providing a clear signature of the parastatistics in the energy spectrum for PBC. Each of these levels will carry an additional degeneracy of $\mbox{dim}\,\mathcal{H}^{\rm fl}_{N,q}$ due to the flavor sector, see Eq.~\eqref{dim2}. The parameters $v$ and $K$ (with $K=1$ in the free case) are known exactly from the Bethe ansatz solution of the XXZ model for arbitrary filling.

\emph{Thermodynamics}---In the thermodynamic limit, the $1/L$-terms due to the boundary conditions become irrelevant. This means that we can start from the open boundary case when considering the $L\to\infty$ limit where the parastatistics only leads to additional degeneracies due to the flavor sector while the occupation number sector is not affected. Returning to our example, this means that the free energy per site is given by ($\beta=1/T$)
\begin{eqnarray}
\label{thermo1}
f&=&-\frac{T}{L}\ln\Tr\, \e^{-\beta H} =-\frac{T}{L}\ln\bigg(\sum_{N=0}^L m^N\sum_{n=1}^{\binom{L}{N}}\e^{-\beta\varepsilon_n}\bigg) \nonumber \\
&=& -\frac{T}{L}\ln\bigg(m^{L/2}\e^{-\beta E_0}\sum_{N=0}^L m^{N-L/2}\sum_{n=1}^{\binom{L}{N}}\e^{-\beta(\varepsilon_n-E_0)} \bigg) \nonumber \\
&=& \frac{E_0}{L}-\frac{\ln m}{2}T-\frac{T}{L}\ln\bigg(\sum_{N,n}\e^{-\beta(\varepsilon_n-E_0-T\ln m(N-\frac{L}{2}))}\bigg) \nonumber \\
&=& \frac{E_0}{L}-\frac{\ln m}{2}T + f^{\rm XXZ}(\mu(T))
\end{eqnarray}
with $f^{\rm XXZ}$ the XXZ free energy per site \cite{Klumper92};  $\mu(T)=T\ln m$ is a temperature-dependent chemical potential where the filling is measured as usual with respect to the half-filled case $N=L/2$. This formula is valid for any temperature and shows the two main effects of the parastatistics: (i) \blue{a zero-temperature entropy density $s_0=-\frac{\partial f}{\partial T}(T\to 0)=\frac{\ln m}{2}$} due to the macroscopic ground-state degeneracy introduced by the flavor sector, and (ii) a chemical potential shift with temperature similar to that of the degenerate Fermi gas. Note that both effects disappear, as expected, for $m=1$,  i.e., the case without flavor degeneracies. \\
{\bf Entropy and temperature-dependent chemical potential:} At low temperatures we can use the conformal field theory result for the XXZ chain and expand to leading order in the (then small) chemical potential, resulting in
\begin{equation}
    \label{thermo2}
    f=\frac{E_0}{L}-\frac{\ln m}{2}T-\frac{\pi c}{6v}T^2-\frac{\chi}{2}(\ln m)^2T^2 +\mathcal{O}(T^3)
\end{equation}
where $\chi=K/(\pi v)$ is the compressibility at half filling. The third term is the universal finite-temperature correction in conformal field theory \cite{Affleck86} while the second and fourth terms are the two signatures of the parastatistics. 

\emph{Conclusions}---In this work we have extended the study of open, non-interacting paraparticle chains to the periodic, interacting setting, \js{thereby establishing the first interacting benchmark systems for R-parastatistics}. We have proven that for flavor-blind Hamiltonians the Hilbert space separates into an occupation and a flavor part and have derived explicit general formulas to count the degeneracies of the eigenspectrum of $H_{\rm occ}$ due to the flavor part both for OBC and PBC. For the PBC case, we have shown, furthermore, that the parastatistics leads to a separation of $H_{\rm occ}$ into flux sectors at fixed particle number $N$. \js{The statistics encoded in the R-algebra thus has \ds{directly} observable spectral and thermodynamic consequences in interacting many-body systems.} As an illustrative example, we considered hardcore paraparticles with $m$ flavors. For this model $H_{\rm occ}$ is the XXZ Hamiltonian, thus allowing an exact determination of the interacting paraparticle spectra and bulk thermodynamics. \blue{The exact solvability of the XXZ chain is useful here for determining parameters such as $E_0$, $v$, and $K$, but the factorization, flux-sector structure, and bosonization description do not rely on integrability. More broadly, our work illustrates how a Hilbert space can factorize into physically distinct sectors; related but complementary ideas of embedding special sectors into larger Hilbert spaces have also been recently explored from the perspective of weak ergodicity breaking} \cite{Katsura25}. For the future, it is interesting to consider also models with non-trivial actions $\rho_n(C)$ on the flavor space. Then, the projector formula still applies but the flavor characters $\chi_n(C^r)$ are no longer simply given by the mode-degeneracies $d_n$.

\acknowledgments
This work was supported by the D-ITP consortium, a program of the Dutch Research Council (NWO) that is funded by the Dutch Ministry of Education, Culture and Science (OCW). J.S.~acknowledges support by NSERC via the Discovery grants program and gratefully acknowledges the hospitality of Utrecht University where part of this work was performed.

\newpage

\onecolumngrid
\begin{center}
\large\bfseries End Matter
\end{center}
\twocolumngrid

\js{
\emph{Appendix A: From Braid Relations to Symmetric-Group Statistics}---In the main part of the paper, we have defined the permuted $R$ matrix via its components $R_{ab}^{cd}$ in the exchange relation \eqref{comm}. As an operator, $R$ acts on $\mathcal{F}\otimes \mathcal{F}$ where $\mathcal{F}$ is the local flavor space. This leads to the relation
\begin{equation}
    \label{EM1}
    R(|a\rangle\otimes |b\rangle) = \sum_{cd} R_{ab}^{cd} (|c\rangle\otimes|d\rangle) \, .
\end{equation}
On the flavor spaces $\mathcal{F}\otimes \mathcal{F}\otimes \mathcal{F}$ we can then define the operators $R_{12}=R\otimes\mathbbm{1}$ and $R_{23}=\mathbbm{1}\otimes R$. In this operator notation, the first two relations below Eq.~\eqref{comm} read
\begin{equation}
    \label{EM2}
    R^2=\mathbbm{1}\, ,\quad  \,R_{12}R_{23}R_{12} = R_{23}R_{12}R_{23}\, .
\end{equation}
\ds{If for a flavor string of length $M$} we define operators $B_i=\mathbbm{1}^{\otimes(i-1)}\otimes R\otimes\blue{\mathbbm{1}^{\otimes(M-i-1)}}$ acting on \blue{$\mathcal{F}^{\otimes M}$} then we see that the constant YBE (second equation in \eqref{EM2}) defines the braid relation
\begin{equation}
    \label{EM3}
    B_i B_{i+1} B_i = B_{i+1}B_i B_{i+1} \, .
\end{equation}
The additional relation $R^2=\mathbbm{1}$ implies that the braid generators $B_i$ fulfill $B_i^2=\mathbbm{1}$. This means that the braid generators are their own inverse, which is akin to forgetting how the strands are braided. Due to this involutive relation, the braid group is reduced to the symmetric \ds{group $S_M$ for $M$ flavor factors.} In particular, the $R$-parastatistics therefore does not allow for anyonic exchange relations. The eigenvalues of $R$ are $\pm 1$, which means that general anyonic phases $\text{e}^{i\varphi}$ are not allowed and that the two-particle flavor space can always be decomposed as $\mathcal{F}\otimes\mathcal{F}=\mathcal{F}_+\oplus\mathcal{F}_-$ where $\mathcal{F}_{\pm}$ denote the $\pm1$ eigenspaces of $R$. Note, however, that this does not mean that $R$-paraparticles always reduce to independent fermions and bosons. Non-trivial constraints can couple flavor sectors and restrict the allowed symmetric-group representations, leading to generalized exclusion rules distinct both from anyonic braiding and from Green's parastatistics, which is based on trilinear rather than quadratic exchange relations.

\emph{Appendix B: Proof of Factorization Theorem}---We sketch a constructive proof of the factorization theorem stated in the main text. 
We assume the quadratic algebra \eqref{comm} and a flavor-blind Hamiltonian of the form \eqref{Ham}, i.e., all terms sum over the internal flavor index.

At each site $i$, we can generate the local Hilbert space from the vacuum by the creation operators $\psi^+_{i,a}$. The local states can be labeled by the eigenvalues of the number operator $n_i=\sum_a \psi^+_{i,a}\psi^-_{i,a}$. The local Hilbert space therefore decomposes as
\begin{equation}
    \label{EM4}
    \mathcal H_i=\bigoplus_{n=0}^{n_{\rm max}} \mathcal H_i^{(n)},
\end{equation}
where $\mathcal H_i^{(n)}$ is the $n$-particle subspace at site $i$. Any state at site $i$ is generated from the vacuum by an ordered product $\psi^+_{i,a_1}\cdots\psi^+_{i,a_n}|0\rangle$ and different orderings are related by the exchange relation \eqref{comm}. This means that the $n$-particle subspace at site $i$ further decomposes into a one-dimensional occupation vector space and a flavor multiplicity space
\begin{equation}
    \label{EM5}
    \mathcal H_i^{(n)} \simeq \mathrm{span}\{|n_i\rangle\} \otimes \mathcal F_n \, .
\end{equation}
with $\mbox{dim}\mathcal F_n = d_n$. The entire Hilbert space therefore decomposes as a direct sum of occupation and multiplicity spaces
\begin{equation}
    \label{EM5b}
    \mathcal H = \bigoplus_{\{n_i\}} \blue{\mathrm{span}\{|\{n_i\}\rangle\}} \otimes \mathcal F(\{n_i\}) \, .
\end{equation}
This establishes the decomposition of the Hilbert space in Eq.~\eqref{Hilbert} of the main text.

The remaining task is to understand how a general flavor-blind Hamiltonian, such as the Hamiltonian \eqref{Ham} in the main text, acts on this Hilbert space. A useful first step is to consider an example with $N=2$ particles on two sites. A general state in the $(1,1)$ sector is then given by
\begin{equation}
    \label{EM6}
    |\Psi\rangle=\sum_{ab}C_{ab}\psi^+_{1a}\psi^+_{2b}|0\rangle \,.
\end{equation}
Now consider the action of the local hopping term $T_{12}=J\sum_c (\psi^+_{1c}\psi^-_{2c}+\psi^+_{2c}\psi^-_{1c})$ on this state
\begin{equation}
    \label{EM7}
    T_{12}|\Psi\rangle = J\sum_{ab} C_{ab} (\psi^+_{2a}\psi^+_{2b}+\sum_{c\gamma} R_{ab}^{c\gamma} \psi^+_{1c}\psi^+_{1\gamma})|0\rangle \, ,
\end{equation}
where we have used the exchange relation \eqref{Para}. The important point is that the flavor coefficients are modified only through the universal linear map $(RC)_{c\gamma}=\sum_{ab} R^{c\gamma}_{ab}C_{ab}$ induced by the exchange algebra \eqref{Para}. 
This map is fixed by the $R$-statistics and does not depend on the Hamiltonian. Thus the hopping term changes the occupation sector, here $(1,1)\to(2,0)+(0,2)$, but introduces no additional flavor dependence beyond that already encoded in the quadratic algebra.

More generally, because all terms in $H$ sum over flavor indices, the Hamiltonian is invariant under unitary rotations of the flavor labels which preserve the quadratic exchange algebra. Consequently, $\mathcal H$ decomposes into invariant flavor sectors $s$ (irreducible representations of this symmetry algebra, which commutes with $H$). Within each sector $s$, $H$ cannot resolve the corresponding flavor multiplicity space and therefore acts trivially on it. Hence, in a basis adapted to this sector decomposition,
\begin{equation}
\label{EM8}
H \simeq \bigoplus_s  H_{\rm occ}^{(s)} \otimes \mathbbm{1}^{(s)}_{\rm fl}\, ,
\end{equation}
where $H_{\rm occ}^{(s)}$ acts on the occupation-number degrees of freedom restricted to sector $s$ and $\mathbbm{1}^{(s)}_{\rm fl}$ acts on the associated multiplicity space. This establishes the factorization theorem claimed in the main text which is the basis for the OBC and PBC degeneracy rules.
}

\newpage

\phantom{O}

\newpage

\onecolumngrid
\begin{center}
\textbf{Supplemental material for ``\blue{Parastatistics in Interacting Periodic Chains Revealed by Peierls Phase Twists and Shifted Conformal Towers}"}\\[3mm]
{Dirk Schuricht}\\
{Institute for Theoretical Physics, Center for Extreme Matter and Emergent Phenomena, Utrecht University, Leuvenlaan 4, 3584 CE Utrecht, The Netherlands}\\[2mm]
{Jesko Sirker}\\
{Department of Physics and Astronomy and Manitoba Quantum Institute, University of Manitoba, Winnipeg, Canada R3T 2N2}\\[5mm]
\end{center}

\noindent
For $F=2$, all solutions of the constant Yang-Baxter equation have been classified by Hietarinta \cite{Hietarinta}, yielding 23 distinct solutions. Here we consider the permuted R-matrix for all these solutions and investigate the additional constraints imposed by the involution condition $R^2=\mathbbm{1}$, which is part of the definition of paraparticle statistics. We also check the additional constraints imposed by unitarity, $R^\dagger R=\mathbbm{1}$, which are needed to obtain Hermitian Hamiltonians. Imposing both conditions, we find that there are only three inequivalent R-matrices for $F=2$. In all of these cases, we constructively obtain the dimensions $d_n$ of the local flavor spaces and also discuss the occupation-flavor factorization for two particularly interesting R-matrices.

\section{General setup}

We work in the second quantized formulation of parastatistics recently introduced in Ref.~\onlinecite{WangHazzard25}. Specifically, we consider a one-dimensional chain of $L$ lattice sites, at each of which we define operators $\psi_{i,a}^\pm$, $i=1,\ldots,L$, where $a=1,\ldots, F$ denotes an internal flavor. The parastatistics is encoded in the commutation relations 
\begin{equation}
\psi_{i,a}^+\psi_{j,b}^+=\sum_{cd} R^{cd}_{ab}\psi_{j,c}^+\psi_{i,d}^+,\quad
\psi_{i,a}^-\psi_{j,b}^-=\sum_{cd} R^{ba}_{dc}\psi_{j,c}^-\psi_{i,d}^-,\quad
\psi^-_{i,a}\psi^+_{j,b} =\sum_{cd} R^{ac}_{bd} \psi^+_{j,c}\psi^-_{i,d}+\delta_{ab}\delta_{ij},
\label{eq:Para}
\end{equation}
where $R_{ab}^{cd}$ are $F^2\times F^2$ matrices. Consistency of the above relations requires 
\begin{equation}
\sum_{\sigma\tau}R_{ab}^{\sigma\tau}R_{\sigma\tau}^{cd}=\delta_a^c\delta_b^d
\label{eq:inverse}
\end{equation} 
as well as 
\begin{equation}
\sum_{\sigma\tau\kappa}R_{ab}^{\sigma\tau}R_{\tau c}^{\kappa u}R_{\sigma\kappa}^{de}=\sum_{\sigma\tau\kappa} R_{bc}^{\sigma\tau}R_{a\sigma}^{d\kappa}R_{\kappa\tau}^{eu}.
\label{eq:cYBE}
\end{equation} 
The latter is known as the (permuted) constant Yang--Baxter equation. The $R$-matrix defined in \eqref{eq:Para} is known  in the literature on integrable systems~\cite{SamajBajnok13} as the permuted $R$-matrix. In the special case $R_{ab}^{cd}=\pm\delta_a^d\delta_b^c$ the relations \eqref{eq:Para} simplify to bosons and fermions with $F$ internal degrees of freedom, while $R_{ab}^{cd}=-\delta_a^c\delta_b^d$ defines example 3 studied in Ref.~\onlinecite{WangHazzard25}. Furthermore, provided the R-matrix satisfies the additional unitarity condition
\begin{equation}
\sum_{\sigma\tau}R_{ab}^{\sigma\tau}(R_{cd}^{\sigma\tau})^*=\delta_{ac}\delta_{bd},
\label{eq:unitarity}
\end{equation}
the operators $\psi_{i,a}^\pm$ satisfy $\psi_{i,a}^+=(\psi_{i,a}^-)^\dagger$ and can be interpreted as usual creation and annihilation operators.

Note that the relations \eqref{eq:inverse}--\eqref{eq:unitarity} are invariant under the global sign change 
\begin{equation}
R_{ab}^{cd}\to -R_{ab}^{cd}.
\label{eq:signchange}
\end{equation}

\section{Hietarinta's solutions}

Hietarinta has classified~\cite{Hietarinta} all solutions $S_{ab}^{cd}$ to the constant Yang--Baxter equation for $F=2$. (We note that Hietarinta uses $R$ for what we denote by $S$ here.) The relation to the permuted R-matrix used in the commutation relations \eqref{eq:Para} is given by $R_{ab}^{cd}=S_{ba}^{cd}$. It is straightforward to check the relations \eqref{eq:inverse}--\eqref{eq:unitarity} for these solutions; we list our findings in Table~\ref{tab:parameters}. 
\begin{table}[ht]
\begin{center}
\begin{tabular}{|c|c|c|c|}
\hline
code & cYBE \eqref{eq:cYBE} satisfied & \eqref{eq:inverse} satisfied (if) & \eqref{eq:unitarity} satisfied (if)\\
\hline
H3.1 & yes & $k^2=s^2=pq=1$ & $|k|^2=|p|^2=|q|^2=|s|^2=1$\\
\hline
H2.1 & yes & $k^4=1$, $k^2=pq$ & $|k|^4=|kp|^2=|kq|^2=1, k^2=pq$\\
H2.2 & yes & $k^4=1$, $k^2=pq$ & $|k|^4=|kp|^2=|kq|^2=1$, $k^2=pq$\\
H2.3 & yes & $k^2=1$, $p+q=p^2+ks=0$ & $|k|^2=1$, $p=q=s=0$\\
H2.4 & yes & no & no \\
\hline
H1.1 & yes & $p^2=q^2$, $4p^4=1$ & $|p|^4=|q|^4$, $4|p|^4=1$ \\
H1.2 & yes & $p=q$, $p^2=1$ & $p=q$, $|p|^2=1$, $k=0$\\
H1.3 & yes & $k^4=1$ & $|k|^4=1$, $p=q=0$\\
H1.4 & yes & $k^2=pq=1$ & $|k|^2=|p|^2=|q|^2=1$\\
H1.5-H1.12 & yes & no & no \\
\hline
H0.1, H0.2 & yes & no & no \\
H0.3 & yes & yes & yes \\
H0.4-H0.6 & yes & no & no \\
\hline
\end{tabular}
\end{center}
\caption{Properties of the permuted R-matrices obtained from Hietarinta's solutions.~\cite{Hietarinta}}
\label{tab:parameters}
\end{table}

In the following, we consider the possible choices in more detail. As it turns out, apart from H1.4 and H0.3, they are all special cases of H3.1. Thus after imposing involution and unitarity there are, for $F=2$, essentially only three inequivalent R-matrices.

\section{Explicit solutions and number of $\textbf{n}$-particle states}
\subsection{Hietarinta's solution H3.1}

Imposing \eqref{eq:inverse} and \eqref{eq:unitarity} the R-matrix H3.1 is explicitly given by
\begin{equation}
\left(R_{ab}^{cd}\right)
=\left(\begin{matrix} 
R_{11}^{11} & R_{11}^{12} & R_{11}^{21} & R_{11}^{22} \\
R_{12}^{11} & R_{12}^{12} & R_{12}^{21} & R_{12}^{22} \\
R_{21}^{11} & R_{21}^{12} & R_{21}^{21} & R_{21}^{22} \\
R_{22}^{11} & R_{22}^{12} & R_{22}^{21} & R_{22}^{22} 
\end{matrix}\right)
=\left(\begin{matrix} 
k & 0 & 0 & 0 \\
0 & 0 & e^{\ii\alpha} & 0 \\
0 & e^{-\ii\alpha} & 0 & 0 \\
0 & 0 & 0 & s
\end{matrix}\right),\quad k,s=\pm1,\;\alpha\in\mathbb{R}.
\label{eq:H3.1}
\end{equation}
We note that the choice $k,s=\pm1$ in total allows for four options with the free parameter $\alpha$ remaining. The sign change \eqref{eq:signchange} does not yield new results.

Specifying to $i=j$ in \eqref{eq:Para} allows to determine the number of states $d_n$ per site $i$ with a given number of particles $n$. Note that $k=\pm 1$ imply bosonic/fermionic exclusion statistics for particles of flavor 1 (and similarly for particles of flavor 2). The off-diagonal elements only fix the mixed exchange phase and do not further reduce the on-site counting. Thus we find 
\begin{eqnarray*}
k=1,\; s=1: && d_{n\ge 0}=n+1,\\
k=\pm 1,\; s=\mp 1: && d_0=1,\,d_{n\ge 1}=2,\\
k=-1,\; s=-1: && d_0=1,\,d_1=2,\,d_2=1,\,d_{n\ge 3}=0.
\end{eqnarray*}
For general $\alpha$, we have the exchange relation between the two flavors $\psi^+_{i1}\psi^+_{i2}=e^{i\alpha}\psi^+_{i2}\psi^+_{i1}$. We note that for $k=s=-1$ and $\alpha=\pi$ we simply obtain fermions with two flavor degrees of freedom. 

As an exemplary case, we consider the factorization of the Hilbert space for $k=-1$ and $s=+1$. If a site is occupied by $n_i=n\geq 1$ particles, then there are two choices: $n$ particles of flavor $s=+1$ which we will denote by $|n,0\rangle$ or $n-1$ particles of flavor $s=+1$ and one particle of flavor $k=-1$ which we will denote by $|n,1\rangle$. The local Hilbert space then factorizes as in Eq.~(3) of the main article
\begin{equation}
\label{fac1}
    \mathcal{H}_i=\bigoplus_{n=0}^\infty \mbox{span}\{|n_i\rangle\}\otimes \mathcal{F}_n
\end{equation}
with $\mbox{dim}\,\mathcal{F}_0=1$ and $\mbox{dim}\,\mathcal{F}_{n\geq 1}=2$ where $\mathcal{F}_{n\geq 1}=\mbox{span}\{|l=0\rangle,|l=1\rangle\}$. This means that for each site we have an occupation label $n=0,1,2,\cdots$ and for each occupied site an additional flavor label $l=0,1$. If we identify $|0\rangle\equiv |0,l\rangle$ then we can again embed the Hilbert space into a larger one, $\mathcal{H}\subset\mathcal{H}^{\rm occ}\otimes\mathcal{H}^{\rm fl}$, just as in the main article. The flavor-blind Hamiltonian (Eq.~(2) in the main text) then again factorizes as in Eq.~(7) of the main article 
\begin{equation}
    H\simeq \bigoplus_{M,q} H^{\rm occ}_{M,q}\otimes \mathbbm{1}_{M,q}
\end{equation}
within the fixed $M$ subspace where $M$ is the number of sites with $n_i\geq 1$. The occupation part $H^{\rm occ}$ is now a bosonic chain with a phase twist $\gamma_q=2\pi q/M$ for periodic boundary conditions instead of an XXZ chain as in the main text. Diagonalizing the spinless bosonic chain (not exactly solvable) gives the eigenenergies and the flavor part then again determines the degeneracies. A general occupied state has the form $|\Psi_{\rm occ}\rangle=\sum_s a_s |\{n_i\}_s\rangle$ and for OBC the degeneracy is $D=2^{\tilde M}$ with
\begin{equation}
    \tilde M=\#\{i:\; n_i^{(s)}\geq 1 \;\forall s \;\mbox{with}\; a_s\neq 0\}.
\end{equation}
This is fully consistent with the general OBC degeneracy rule in the main text. For PBC, one has to similarly determine the degeneracy of each flux sector characterized by the integer $q$. In particular, the dimension of the flavor space for fixed $M$ is given by 
\begin{equation}
\label{fac2}
    \mbox{dim}\,\mathcal{H}^{\rm fl}_{M,q} =\frac{1}{M}\sum_{r=0}^{M-1}e^{-2\pi i qr/M} 2^{\mbox{gcd}(M,r)} = \frac{1}{M}\sum_{d|M} 2^d \,\mathcal{R}_{M/d}(q)
\end{equation}
which is Eq.~(9) of the main text with $N\to M$ and $m=2$. This formula is fully consistent with the general PBC degeneracy rule, Eq.~(5) of the main text. 

\subsection{Hietarinta's solution H2.1}

\begin{equation}
\left(R_{ab}^{cd}\right)
=\left(\begin{matrix} 
R_{11}^{11} & R_{11}^{12} & R_{11}^{21} & R_{11}^{22} \\
R_{12}^{11} & R_{12}^{12} & R_{12}^{21} & R_{12}^{22} \\
R_{21}^{11} & R_{21}^{12} & R_{21}^{21} & R_{21}^{22} \\
R_{22}^{11} & R_{22}^{12} & R_{22}^{21} & R_{22}^{22} 
\end{matrix}\right)
=\left(\begin{matrix} 
\pm 1 & 0 & 0 & 0 \\
0 & 0 & e^{\ii\alpha} & 0 \\
0 & e^{-\ii\alpha} & 0 & 0 \\
0 & 0 & 0 & \pm 1
\end{matrix}\right),\quad \alpha\in\mathbb{R}.
\label{eq:H2.1}
\end{equation}
This is a special case of H3.1 with $s=k$.

\subsection{Hietarinta's solution H2.2}

\begin{equation}
\left(R_{ab}^{cd}\right)
=\left(\begin{matrix} 
R_{11}^{11} & R_{11}^{12} & R_{11}^{21} & R_{11}^{22} \\
R_{12}^{11} & R_{12}^{12} & R_{12}^{21} & R_{12}^{22} \\
R_{21}^{11} & R_{21}^{12} & R_{21}^{21} & R_{21}^{22} \\
R_{22}^{11} & R_{22}^{12} & R_{22}^{21} & R_{22}^{22} 
\end{matrix}\right)
=\left(\begin{matrix} 
\pm 1 & 0 & 0 & 0 \\
0 & 0 & e^{\ii\alpha} & 0 \\
0 & e^{-\ii\alpha} & 0 & 0 \\
0 & 0 & 0 & \mp 1
\end{matrix}\right),\quad \alpha\in\mathbb{R}.
\label{eq:H2.2}
\end{equation}
This is a special case of H3.1 with $s=-k$.

\subsection{Hietarinta's solutions H2.3 and H1.3}

\begin{equation}
\left(R_{ab}^{cd}\right)
=\left(\begin{matrix} 
R_{11}^{11} & R_{11}^{12} & R_{11}^{21} & R_{11}^{22} \\
R_{12}^{11} & R_{12}^{12} & R_{12}^{21} & R_{12}^{22} \\
R_{21}^{11} & R_{21}^{12} & R_{21}^{21} & R_{21}^{22} \\
R_{22}^{11} & R_{22}^{12} & R_{22}^{21} & R_{22}^{22} 
\end{matrix}\right)
=\left(\begin{matrix} 
\pm 1 & 0 & 0 & 0 \\
0 & 0 & \pm 1 & 0 \\
0 & \pm 1 & 0 & 0 \\
0 & 0 & 0 & \pm 1
\end{matrix}\right).
\label{eq:H2.3}
\end{equation}
These are just two flavors of bosons (upper sign) or fermions (lower sign). It is a special case of H3.1 with $s=k$ and $\alpha=0$ or $\alpha=\pi$. 

The lower sign is identical to example 1 in Ref.~\onlinecite{WangHazzard25}. It is also identical to example 2 since we consider $F=2$ only. Furthermore, choosing $M=\ii\pi\sigma_y/2$ in example 4 we get $\lambda=\xi=-\ii\sigma_y$ and thus again H2.3.

\subsection{Hietarinta's solution H1.1}

\begin{equation}
\left(R_{ab}^{cd}\right)
=\left(\begin{matrix} 
R_{11}^{11} & R_{11}^{12} & R_{11}^{21} & R_{11}^{22} \\
R_{12}^{11} & R_{12}^{12} & R_{12}^{21} & R_{12}^{22} \\
R_{21}^{11} & R_{21}^{12} & R_{21}^{21} & R_{21}^{22} \\
R_{22}^{11} & R_{22}^{12} & R_{22}^{21} & R_{22}^{22} 
\end{matrix}\right)
=\left(\begin{matrix} 
\pm 1 & 0 & 0 & 0 \\
0 & 0 & 1 & 0 \\
0 & 1 & 0 & 0 \\
0 & 0 & 0 & \mp 1
\end{matrix}\right).
\label{eq:H1.1}
\end{equation}
This is a special case of H3.1 with $s=-k$ and $\alpha=0$. 

\subsection{Hietarinta's solution H1.2}

\begin{equation}
\left(R_{ab}^{cd}\right)
=\left(\begin{matrix} 
R_{11}^{11} & R_{11}^{12} & R_{11}^{21} & R_{11}^{22} \\
R_{12}^{11} & R_{12}^{12} & R_{12}^{21} & R_{12}^{22} \\
R_{21}^{11} & R_{21}^{12} & R_{21}^{21} & R_{21}^{22} \\
R_{22}^{11} & R_{22}^{12} & R_{22}^{21} & R_{22}^{22} 
\end{matrix}\right)
=\left(\begin{matrix} 
\pm 1 & 0 & 0 & 0 \\
0 & 0 & \pm 1 & 0 \\
0 & \pm 1 & 0 & 0 \\
0 & 0 & 0 & \mp 1
\end{matrix}\right).
\label{eq:H1.2}
\end{equation}
This is a special case of H3.1 with $k=-s=1$ and $\alpha=0$ or $k=-s=-1$ and $\alpha=\pi$. 

\subsection{Hietarinta's solution H1.4}

\begin{equation}
\left(R_{ab}^{cd}\right)
=\left(\begin{matrix} 
R_{11}^{11} & R_{11}^{12} & R_{11}^{21} & R_{11}^{22} \\
R_{12}^{11} & R_{12}^{12} & R_{12}^{21} & R_{12}^{22} \\
R_{21}^{11} & R_{21}^{12} & R_{21}^{21} & R_{21}^{22} \\
R_{22}^{11} & R_{22}^{12} & R_{22}^{21} & R_{22}^{22} 
\end{matrix}\right)
=\left(\begin{matrix} 
0 & 0 & 0 & e^{\ii\alpha} \\
0 & \pm 1 & 0 & 0 \\
0 & 0 & \pm 1 & 0 \\
e^{-\ii\alpha} & 0 & 0 & 0
\end{matrix}\right),\quad \alpha\in\mathbb{R}.
\label{eq:H1.4}
\end{equation}
The upper right element implies $\psi_{i1}^+\psi_{i1}^+=e^{\ii\alpha}\psi_{i2}^+\psi_{i2}^+$, i.e., states on doubly occupied sites are not independent. In addition, the choice of the lower sign implies $\psi_{i1}^+\psi_{i2}^+=0$, ie, the site $i$ cannot be occupied by two particles of different flavor. Hence we find 
\begin{equation}
d_0=1,\,d_{n\ge 1}=2\;\mbox{(upper sign)}\quad\text{or}\quad d_0=1,\,d_1=2,\,d_{n\ge 2}=1 \;\mbox{(lower sign)}.
\end{equation}

The lower sign is an interesting example for a non-trivial local constraint leading to new exclusion statistics. This case can be interpreted as one boson plus one fermion with a hardcore constraint between the two species. A single occupied site can therefore contain either a boson or a fermion and thus carries a flavor degree of freedom, while all higher occupancies are solely made out of bosons and therefore carry no flavor degree of freedom. We can again factorize the local Hilbert space as in Eq.~\eqref{fac1} but now $\mbox{dim}\,\mathcal{F}_0=1$, $\mbox{dim}\,\mathcal{F}_1=2$, and $\mbox{dim}\,\mathcal{F}_{n\geq 2}=1$. For the factorization of the Hamiltonian this means that we have again a bosonic occupation number chain and the degeneracies for OBC are $D=2^{\tilde M}$ where now
\begin{equation}
    \tilde M=\#\{i:\; n_i^{(s)}= 1 \;\forall s \;\mbox{with}\; a_s\neq 0\}.
\end{equation}
For PBC we have to define similarly $M$ as the number of sites which are singly occupied, $n_i=1$. Then the degeneracy formula \eqref{fac2} is again applicable for sectors with $M$ fixed. These results are again fully consistent with the general formulas in the main text.

\subsection{Hietarinta's solution H0.3}

Keeping in mind the additional freedom of the sign change \eqref{eq:signchange}, we have
\begin{equation}
\left(R_{ab}^{cd}\right)
=\left(\begin{matrix} 
R_{11}^{11} & R_{11}^{12} & R_{11}^{21} & R_{11}^{22} \\
R_{12}^{11} & R_{12}^{12} & R_{12}^{21} & R_{12}^{22} \\
R_{21}^{11} & R_{21}^{12} & R_{21}^{21} & R_{21}^{22} \\
R_{22}^{11} & R_{22}^{12} & R_{22}^{21} & R_{22}^{22} 
\end{matrix}\right)
=\left(\begin{matrix} 
\pm 1 & 0 & 0 & 0 \\
0 & \pm 1 & 0 & 0 \\
0 & 0 & \pm 1 & 0 \\
0 & 0 & 0 & \pm 1
\end{matrix}\right).
\label{eq:H0.3}
\end{equation}
For the upper sign there is no constraint on the occupations and we simply obtain $d_{n\ge 0}=n+1$. In the case of the lower sign, however, we get 
\begin{equation}
R_{ab}^{cd}=-\delta_a^c\delta_b^d\quad\rightarrow\quad d_0=1,\,d_1=2,\,d_{n\ge 2}=0.
\end{equation}
This is example 3 of Ref.~\onlinecite{WangHazzard25}.

\section{Local occupation numbers $d_n$}
The above analysis shows that there are five different possibilities for the local occupation numbers $d_n$, which can always be interpreted as local bosonic or fermionic exchange channels (because of the involution condition $R^2=\mathbbm{1}$), but with local constraints leading to novel exclusion statistics different from ordinary bosons or fermions. These five $\{d_n\}$ combinations arise from the three inequivalent solutions for the $R$-matrix (after imposing the two conditions $R^2=\mathbbm{1}$ and $R^\dagger R=\mathbbm{1}$) via the remaining freedom to choose different signs in the $R$ matrix. The five possibilities are:
\begin{eqnarray*}
\text{(i)} && d_{n\ge 0}=n+1\quad\to\quad\text{two bosons},\\
\text{(ii)} && d_0=1,\,d_{n\ge 1}=2\quad\to\quad\text{boson and fermion},\\
\text{(iii)} && d_0=1,\,d_1=2,\,d_2=1,\,d_{n\ge 3}=0\quad\to\quad\text{two fermions},\\
\text{(iv)} && d_0=1,\,d_1=2,\,d_{n\ge 2}=1\quad\to\quad\text{boson and fermion with hard-core constraint},\\
\text{(v)} && d_0=1,\,d_1=2,\,d_{n\ge 2}=0\quad\to\quad\text{two fermions with hard-core constraint}.
\end{eqnarray*}

\end{document}